\documentclass[10pt,floatfix]{paper}

\usepackage{bm}
\usepackage{color}
\usepackage{subfig}
\usepackage[rightcaption]{sidecap}
\usepackage{todonotes}
\usepackage{graphicx}
\usepackage{subfig}
\usepackage{caption}
\usepackage{amssymb,amsfonts,amsmath,empheq}
\usepackage{authblk}

\usepackage[margin=1.in]{geometry}
\usepackage{pdflscape}
\usepackage{afterpage}

\usepackage[makeroom]{cancel}
\newcommand{\dbar}{{\mkern0.75mu\mathchar '26\mkern -9.75mu {\rm d}}}
\def\d{\mathrm{d}}
\def\e{\mathrm{e}}
\begin{document}

\title{On the nonequilibrium entropy of large and small systems}

\author{
Sheldon Goldstein\thanks{Department of Mathematics, Rutgers University, Hill Center, 110 Frelinghuysen Road, Piscataway, NJ 08854-8019, USA. E-mail: oldstein@math.rutgers.edu}, 
{ David A. Huse}\thanks{Department of Physics, Princeton University, Jadwin Hall, Washington Road, Princeton, NJ 08544-0708, USA. E-mail: huse@princeton.edu},
{ Joel L. Lebowitz}\thanks{Departments of Mathematics and Physics, Rutgers University, Hill Center, 110 Frelinghuysen Road, Piscataway, NJ 08854-8019, USA. E-mail: lebowitz@math.rutgers.edu},
{ Pablo Sartori} \thanks{The Center for Studies in Physics and Biology and Laboratory of Living Matter, Rockefeller University, New York, NY 10065, USA. E-mail: psartori@rockefeller.edu}
}

\maketitle

\begin{abstract}
Thermodynamics makes definite predictions about the thermal behavior of macroscopic systems in and out of equilibrium. Statistical mechanics aims to derive this behavior from the dynamics and statistics of the atoms and molecules making up these systems. A key element in this derivation is the large number of microscopic degrees of freedom of macroscopic systems. Therefore, the extension of thermodynamic concepts, such as entropy, to small (nano) systems raises many questions. Here we shall reexamine various definitions of entropy for nonequilibrium systems, large and small. These include thermodynamic (hydrodynamic), Boltzmann, and Gibbs-Shannon entropies. We shall argue that, despite its common use, the last is not an appropriate physical entropy for such systems, either isolated or in contact with thermal reservoirs: physical entropies should depend on the microstate of the system, not on a subjective probability distribution.  To square this point of view with experimental results of Bechhoefer we shall argue  that the Gibbs-Shannon entropy of a nano particle in a  thermal fluid should be interpreted as the Boltzmann entropy of a  dilute gas of Brownian particles in the fluid.

\end{abstract}

\section{Introduction}

The role of probability in the statistical mechanical analysis of the thermal behavior of individual physical systems is subtle. Indeed, it has frequently been a source of confusion and controversy: note e.g. the conflict between Boltzmann and Zermelo about the H-theorem \cite{klein1973development}. A crucial ingredient in the statistical mechanical analysis of this problem by Maxwell, Thomson, Boltzmann, Gibbs and Einstein is the ``law of large numbers'', which permits ``almost sure'' predictions, i.e. with probability approaching 1, when the number of quasi-independent entities in the system become very large. This is clearly the case for macroscopic systems (MS), which contain a large number of atoms or molecules, to which statistical mechanics was historically restricted. Thus the microcanonical ensemble and the other equilibrium Gibbs ensembles make definite predictions for equilibrium MS. We can therefore speak of the ``typical'' behavior of such a system\footnote{We are taking for granted here an assumed underlying (approximately) equal a priori probability of different microstates for a specified macrostate.}. This restriction to MS was historically natural, since the notions of heat, entropy and the second law were all developed in the nineteenth century for such systems.  The subsequent development of statistical mechanics had as its aim to describe and explain microscopically the observed thermal phenomena in such MS. It therefore also considered only systems  consisting of very large numbers of particles.

In going beyond equilibrium, where the theory is fundamentally complete, this disparity in sizes between microscopic and macroscopic also plays a critical role.  It forms the basis of the explanation by Boltzmann, Maxwell and Thomson of how time-asymmetric behavior, as expressed for example by the second law of thermodynamics, can originate from time-symmetric microscopic laws \cite{penrose2005foundations, lebowitz2007time}.  In particular, time-asymmetric macroscopic equations like the heat equation or the Navier-Stokes equations, as well as the mesoscopic Boltzmann equation (to which Zermelo objected), can be seen as being expressions of the law of large numbers, valid in the limit of particle number $N\to \infty$ \cite{goldstein2004boltzmann}. Unfortunately, a rigorous mathematical derivation of such equations from time-symmetric microscopic dynamics is still beyond our reach for realistic systems. In fact, the only cases for which hydrodynamic-type equations have been derived rigorously are systems with bulk stochastic interactions, like lattice gases \cite{kipnis2013scaling, giacomin1998deterministic, lebowitz1988microscopic}. Therefore,  there are still many open problems for nonequilibrium {MS}. 

The reliance on the law of large numbers raises the issue of understanding the thermal behavior of nanosystems (NS), in which there is currently much interest.  This interest is fueled by technological advances that make such systems experimentally accessible.  Nanosystems can be well isolated from their environment, or can be in contact with reservoirs.  Here we will focus on the latter case, 
an example being the recent work of {Bechhoefer et al }\cite{gavrilov2017direct}, of a nanoparticle immersed in a fluid (a talk by Bechhoefer triggered this work). Recent work on such NS goes under the name of ``stochastic thermodynamics'', see \cite{ciliberto2017experiments} and other articles in that issue. Stochastic thermodynamics, as the name indicates, takes explicit account of the stochastic modeling of the effective interactions between the small system, such as a nanoparticle or a polymer, and the equilibrium thermal reservoir that it is in contact with, usually a macroscopic fluid. There is also much current interest in isolated quantum systems having only a few degrees of freedom \cite{kaufman2016quantum}, but we shall not consider these here.


The consideration of thermal properties of NS raises the question of whether and how the thermodynamic and statistical mechanics formalism can be extended to systems with a small number of degrees of freedom. In the stochastic thermodynamic extensions the Gibbs-Shannon (GS) entropy (defined in (\ref{eq:SGS})) plays a central role. This raises the questions: does the GS entropy of a probability measure $\mu$, when $\mu$ is not a Gibbs measure for an equilibrium macroscopic system, have physical meaning? And when it does, what is that meaning? This entropy has some very nice mathematical properties and it is very alluring to consider it, as is generally done in the stochastic thermodynamic literature, as the ``proper entropy'' of a nonequilibrium system. We shall argue against this for MS. On the other hand the experiments of Bechhoefer, mentioned earlier, actually have measured this quantity, more or less directly, for a nanoparticle immersed in a liquid. We shall discuss our interpretation of these experiments at the end of this note.  Let us however start from the beginning and first consider the statistical mechanical entropy for isolated MS. We shall then consider both MS and NS in contact with thermal baths.

\section{Thermodynamic and Boltzmann entropy of a MS}

The discovery by Clausius of the existence of an entropy function $S(E,V,N)$ for  equilibrium macroscopic systems (with  energy $E$, particle number $N$ and volume $V$), and its central {role} in the time asymmetric evolution of the world, as expressed by the second law, is one of the key events of nineteenth century science, c.f. \cite{brush1967science} and \cite{callen1998thermodynamics}. This discovery raised immediately the question of how to define $S$ as a function of the microstate ${X} = (\vec{r}_1,\vec{p}_1,\ldots,\vec{r}_N,\vec{p}_N)$ of the particles composing the system, where $\vec{r}_i\in V$ is the position and $\vec{p}_i\in \mathbb{R}^3$ the momentum of the $i$th particle. The problem was compounded by the fact that the time evolution $X(t)$ in the phase space $\Gamma$, as given by the Hamiltonian $H(X)$, is time symmetric, c.f. \cite{lebowitz2007time}.

The answer arrived at by Boltzmann was to identify $S$ for a MS with 
\begin{align}
S_{\rm B}(X)=\log|\Gamma(M(X))|\quad.
\end{align}
Here $X$ is a phase point (microstate) in the energy shell, $E \le H(X)\le E+\Delta E$ and $M(X)$ is the macrostate of the system. This macrostate is defined, e.g., by dividing $V$ into $\mathcal{N}$ cells $\omega_\alpha$, $\alpha=1,\ldots\mathcal{N}$, with $1\ll \mathcal{N}\ll N$, and then specifying the number, energy, and total momentum of the particles in each $\omega_\alpha$ with a certain tolerance. $|\Gamma(M)|$ is the Liouville (Lebesgue) volume of the phase space region $\Gamma(M)$ containing all microstates $X$ belonging to the macrostate $M$, c.f. \cite{penrose2005foundations, lebowitz2007time} and also section \ref{sec:bg} of this work.  (For classical systems there is an arbitrary overall additive constant in the entropy coming from the unit of phase space volume, but this has no impact on what we discuss here, so will be ignored.)  For a macroscopic system there is a special macrostate $M_{\rm eq}$, corresponding to equilibrium, such that $\Gamma(M_{\rm eq})$ covers almost the whole surface of energy $E$. That is, $|\Gamma(M_{\rm eq})|\sim|\Gamma_E|$, the volume of the energy shell between $E$ and $E+\Delta E$. This definition of $S_{\rm B}(X)$ assigns an entropy even to microstates $X$ which do not behave at all as expected from the second law, as discussed below after (\ref{eq:2lawbol}). 

The Boltzmann entropy $S_{\rm B}(M_{\rm eq})$ agrees, to leading order in the size of the system, with the experimentally determined equilibrium Clausius entropy $S$. This was shown for a dilute gas by Boltzmann and for general systems by Gibbs. To calculate this entropy Gibbs (and Boltzmann) introduced the microcanonical ensemble $\mu_{\rm m}$, as the uniform probability density in the energy shell $E\le H(X)\le E+\Delta E$ for describing the almost sure properties of equilibrium MS with energy $E$. They naturally equated the precisely defined logarithm of $|\Gamma_E|$, the volume of the energy shell which for MS is very close to $|\Gamma(M_{\rm eq})|$, with the entropy $S(E)$.

{\bf Time-evolution:} The time evolution of the microstate $X(t)$ will  induce a time evolution of the macrostate $M$. Boltzmann then argued that $S_{\rm B}(X(t))$ will, for a typical $X$ in $\Gamma(M)$, evolve in time according to the second law, i.e.
\begin{align}
\frac{\d S_{\rm B}(X(t))}{\d t}\ge0\quad,\quad t>0\label{eq:2lawbol}
\end{align}
see \cite{goldstein2004boltzmann} and references there. This can be proven when one assumes that $M(t)$ evolves under an autonomous macroscopic equation, e.g. the Navier-Stokes or diffusion equation, but the rigorous derivations of these equations from the microscopic dynamics  is not available {at the present time}.

\underline{Nota Bene}: Eq.~(\ref{eq:2lawbol})  can only be true for typical microstates $X$, i.e. for the overwhelming majority of the $X$'s  with respect to Liouville measure in $\Gamma(M)$: there are special microstates for which it is definitely false.  An example of such a special state can be obtained by starting in a typical {low-entropy} state at some time in the past, and then evolving that state in time to a higher-entropy state in the present, followed by exactly reversing all velocities.

{\bf Hydrodynamic time-evolution:} We now describe a class of nonequilibrium systems for which the time evolution of the  Boltzmann entropy is given by  hydrodynamic equations and satisfies the second law. Consider a system in a macrostate $M\neq M_{\rm eq}$ for which one can define a ``smooth'' energy and mass density profile $e(\vec{r})$ and $n(\vec{r})$, where $\vec{r}\in V$ denotes different spatial points of the system. For such systems $S_{\rm B}(X)$  coincides to leading order with $S_{\rm h}(\{e(\vec{r}),n(\vec{r})\})$, the hydrodynamic entropy of systems in local thermal equilibrium (LTE) given by \cite{de2013non}
\begin{align}\label{ent:hydro}
S_{\rm h }(\{e(\vec{r}),n(\vec{r})\})=\int _V s(e(\vec{r}),n(\vec r))\d \vec{r}\quad,
\end{align}
where $s(e,n)$ is the equilibrium entropy per unit volume, $s=S/|V|$ (in the thermodynamic limit). [We have assumed for simplicity that the local velocity $ u(\vec{r})$ is zero, otherwise $e(\vec{r})\to [e(\vec{r})-\frac12 n(\vec{r}) |u(\vec{r})|^2]$.  Note that $S_{\rm h}$ coincides with the equilibrium entropy $S(E,V,N)$ when $e$ and $n$ are independent of $\vec{r}$.


As an example of the hydrodynamic time evolution of $S_{\rm h}$, consider a system in LTE with a temperature profile $T(\vec{r},t)$. Starting then with the general equation for the time evolution of the entropy density of a system in LTE 
\begin{align}
\frac{\partial s(\vec{r},t)}{\partial t} = - \frac{\text{div } j(\vec{r},t)}{T(\vec{r}, t)} =- \text{div } (j/T) + j \cdot \nabla \left(\frac{1}{T}\right)\quad, \label{eq:entlte}
\end{align}
where $j(\vec{r},t)$ is the energy flux vector, we get
\begin{align}\label{eq:boltentbal}
\frac{\d S_{\rm h}}{\d t} = -\int_Q \frac{j(\vec{q},t)}{T(\vec{q})} \cdotp\d \vec{q} +\int_V j(\vec{r}, t)\cdot\nabla  \left(\frac{1}{T(\vec{r},t)}\right) \d\vec{r}   \quad.
\end{align}
In (\ref{eq:boltentbal}) $Q$ is the surface of $V$ and $\d\vec{q}$ is the (outward directed) surface area element. The flux integrand vanishes on the parts of the surface which are insulated, the whole surface if the system is isolated. There will, however, be a contribution from the parts of the surface which are held at specified temperatures $T(\vec{q})$ by external reservoirs. The integral over $Q$ can be identified with the entropy production in the thermal reservoirs, ${\dbar  S_{\rm r}}/{\d t}$, which maintain the temperature $T(\vec{q})$. (Since we have idealized the reservoirs as infinite systems with fixed temperatures, their entropy is formally infinite, but the rate of change in their entropies is finite.) The second term in (\ref{eq:boltentbal}) corresponds to the hydrodynamic or Boltzmann entropy change in the bulk of the MS,
\begin{align}
\sigma_{\rm B}(t) = \int_V j(\vec{r}, t) \cdot \nabla \left(\frac{1}{T (\vec{r}, t)}\right) \d\vec{r}\ge0\quad,\label{eq:boltepr}
\end{align}
due to local ``dissipation''. The integrand is in fact everywhere non-negative, an expression of the second law: the component of the energy flux parallel to the temperature gradient cannot be directed from `cold' towards `hot'.


\section{The Gibbs-Shannon entropy}

The  entropy of the micro-canonical ensemble, $S(E)=\log|\Gamma_E|$, can also be written as
\begin{align}
S(E)=S_{\rm G} (\mu_{\rm m})= -\int \mu_{\rm m}(X)\log\mu_{\rm m}(X) \d X\quad.\label{eq:SG}
\end{align}
Using Legendre transforms, Gibbs showed that if one considers the canonical ensemble with probability density $\mu_\beta$ given by
\begin{align}
\mu_{\beta}=Z^{-1}\exp\left[-\beta H(X)\right]\quad,
\end{align}
with $\beta=1/T$,  then $S_{\rm G}(\mu_\beta)$ also gives the equilibrium entropy of a MS as a function of temperature. The same is true for the grand-canonical ensemble and other equilibrium ensembles. They all agree to leading order in the size of the system.

It is a natural step to extend this notion of entropy to general probability measures with densities $\mu(X,t)$ which depend on $X$ and $t$ as 
\begin{align}
S_{\rm G} (\mu)= -\int \mu(X,t)\log\mu(X,t) \d X\quad.\label{eq:SGS}
\end{align}
The quantity $S_{\rm G}(\mu)$ is the Shannon entropy of an arbitrary measure $\mu$ on a space $\Omega$ (relative to the measure $\d X$).  It plays a central role in information theory as developed by Shannon \cite{shannon1948mathematical}. However, as is well known, for an isolated physical system evolving under Hamiltonian dynamics, $\mu$ changes in time according to the Liouville equation, ${\partial \mu}/{\partial t} = - \{\mu, H\}$, and the GS entropy $S_{\rm G}(\mu(t))$ does not change at all.  Thus $S_{\rm G}(\mu(t))$ cannot be identified with the thermodynamic or hydrodynamic entropy $S_{\rm h}$ of an isolated macroscopic system which is not in global thermal equilibrium, even if it is in local thermal equilibrium, a situation in which $S_{\rm h}$ is unambiguous. (This was already noted by Gibbs and discussed by P. and T. Ehrenfest in their 1916 article \cite{ehrenfest2002conceptual}). This raises the question of what is the physical meaning of $S_{\rm G}(\mu)$ for any system for which $\mu$ is not an equilibrium Gibbs measure of a MS.

The behavior of the GS entropy associated with a measure $\mu$ is very different when the system is in contact with stochastic reservoirs. As will be seen below, the rate of change of $S_{\rm G}$ is no longer zero and is related to the thermodynamic entropy change in the reservoirs. We shall discuss the physical significance, if any, of this later, after we introduce the mathematical formalism to describe such systems.

\section{Model of system in contact with thermal reservoirs}

The formalism we shall  use was developed by Bergmann and Lebowitz \cite{bergmann1955new, lebowitz1957irreversible} who studied the dynamics of a system evolving under the combined action of its own Hamiltonian $H(X)$ and of $n$ thermal reservoirs at different temperatures (and chemical potentials). These reservoirs were thought of as being infinite and acting at the boundaries of the MS. To simplify matters the interaction with the reservoirs was idealized as being of the collision type: when a collision occurs the phase point of the system, $X$, jumps to $X'$, while the reservoir particle goes off to infinity, never to be seen again.  The system thus sees an ever fresh stream of reservoir particles with a Maxwellian distribution, at the temperature $T_\alpha = \beta^{-1}_\alpha$ of that reservoir, $\alpha = 1, \cdots, n$. The time evolution of the system will thus be given by a continuous time Markov process.

Denoting $K_\alpha (X,X') \d X$ the transition rate from the phase point $X'$ to the phase space volume $\d X$ around $X$ due to collisions with reservoir $\alpha$, yields the following stochastic Liouville master equation for the probability density $\mu(X, t)$,
\begin{align}
\frac{\partial \mu(X, t)}{\partial t} + \{\mu, H\} = \sum_{\alpha=1}^n\int \left[K_\alpha (X,X') \mu (X', t) -K_\alpha(X',X) \mu (X,t)\right] \d X'\quad,\label{eq:liouville}
\end{align}
where  $\{\mu, H\}$ is the usual Poisson bracket describing the deterministic Hamiltonian evolution of the isolated system.

Using the time reversibility of the collision dynamics yields a condition for each $\alpha$,
\begin{align}
K_\alpha (X,X') = e^{\beta_\alpha H(X')} L_\alpha (X,X')\label{eq:db}
\end{align}
with  $L_\alpha (X,X') = L_\alpha (\overline{X'}, \overline{X})$, where $\overline{X}$ corresponds to reversal of the velocity coordinates of $X$. Some further simplifications give $L_\alpha(\overline{X'}, \overline{X}) = L_\alpha (X',X)$, so that $L_\alpha({X}, {X}') = L_\alpha (X',X)$, corresponding to ``detailed balance'' for {each reservoir, i.e.}
\begin{align}
K_\alpha(X,X')/K_\alpha(X',X)=\exp[-\beta_\alpha(H(X)-H(X'))]\quad.\label{eq:detbal}
\end{align}

It was proven in \cite{bergmann1955new}, under quite general conditions on $L_\alpha (X,X')$ that, as $t\to \infty$, a system started in some arbitrary initial $\mu(X,0)$ will approach a stationary state 
\begin{align}
\lim_{t\to \infty} \mu (X, t) = \mu_{\rm s}(X)\quad. 
\end{align}
This state is unique and is absolutely continuous with respect to Liouville measure.  When there is only one reservoir at reciprocal temperature, $\beta_\alpha$, then clearly
\begin{align}
\mu_{\rm s}(X) = \mu_\alpha(X) \equiv Z^{-1} \exp \left[ - \beta_\alpha H(X)\right]
\end{align}
is the unique stationary state.  When the temperatures $\beta_\alpha^{-1}$ are different $\mu_{\rm s}$ will be a nonequilibrium stationary state (NESS), for which the dynamics do not satisfy detailed balance.  It was further shown that this NESS will satisfy the Onsager reciprocal relations when all $\beta_\alpha$ are close to some $\overline{\beta}$, as well as a generalized Kubo relation in the presence of an external field.

\section{Time evolution of Gibbs entropy for a system in contact with thermal reservoirs}
\label{sec:timsgibbs}

For a closed system, given the phase point $X(t_0)$, $X(t)$ is determined for all $t$. The only randomness expressed in $\mu(X,t)$ for a closed system is that introduced initially, which could be due to ignorance. There is therefore no intrinsic physical significance to $\mu(X,t)$ for an isolated system. On the other hand when the system is in contact with a stochastic reservoir then $X(t)$ is no longer determined by $X(t_0)$, and $\mu(X,t)$ acquires some ``objective'' meaning. The GS entropy now evolves in time in a non trivial way, which can be calculated from (\ref{eq:liouville}). It consists of two contributions,
\begin{align}
\frac{\d}{\d t} S_{\rm G}(\mu) = \sum_{\alpha=1}^n J_\alpha(t) / T_\alpha +\sigma_{\rm G} (t)\quad.\label{eq:entflux}
\end{align}
In the first contribution $J_\alpha$ is the average energy flux {\it from} the $\alpha$th reservoir into the system, that is
\begin{align}
J_\alpha(t) =\int \mu (X, t) \int K_\alpha (X', X) \left[H(X') - H(X)\right] \d X'  \d X \quad,\label{eq:enerflux}
\end{align}
with $ \sum_{\alpha=1}^n J_\alpha(t)=\frac{\d}{\d t} \int H(X)\mu(X,t) \d X$. The second contribution in (\ref{eq:entflux}) is  
\begin{align}
\sigma_{\rm G} (t) =\frac12\sum_{\alpha=1}^n \int\int L_\alpha (X,X') \left[\nu_\alpha (X,t) - \nu_\alpha (X', t)\right] \log \left[\frac{\nu_\alpha (X, t)}{\nu_\alpha (X', t)}\right] \d X \d X' \geq 0\quad,\label{eq:gibbstot}
\end{align}
where 
\begin{align}
\nu_\alpha (X,t) = \mu(X,t) \exp \left[\beta_\alpha H(X)\right]\quad.\label{eq:nu}
\end{align}

Equation (\ref{eq:entflux}) can be rewritten in the suggestive form
\begin{align}
\sigma_{\rm G} (t)= \frac{\dbar S_{\rm r}}{\d t} + \frac{\d S_{\rm G}}{\d t} \geq 0\quad,\label{eq:suggestive}
\end{align}
where we have written {${\dbar S_{\rm r}}/{\d t} = \sum_\alpha {\dbar S_\alpha}/{\d t}$} and   
\begin{align}
\frac{\dbar S_\alpha}{\d t} =  - J_\alpha / T_\alpha 
\end{align}
is the rate of change of the entropy of the $\alpha$th reservoir caused by the energy (heat) flow $-J_\alpha$ {\it into} that reservoir. The equation (\ref{eq:suggestive}) is reminiscent of the second law, and has therefore prompted the interpretation of $S_{\rm G}(\mu)$ for systems in contact with thermal reservoirs as a physical entropy, despite the fact that it is not so for an isolated system and is not specified by the microstate of the system, c.f. section \ref{sec:last}.

We want to argue however that (\ref{eq:suggestive}) does not justify the interpretation of $S_{\rm G}$ as the physical entropy of an open nonequilibrium system unless it agrees, at least to leading order, with $S_{\rm B}$. In fact for a MS in contact with reservoirs at its surface all the entropy production $\sigma_{\rm G}$ is caused, as can be seen from (\ref{eq:gibbstot}), by the stochastic interactions at its surface. This is in contrast to the entropy production $\sigma_{\rm B}$, given in (\ref{eq:boltepr}), which is due to the chaotic microscopic dynamics in the bulk of the system, as it should be from a physical point of view.

We note further that, as is well known, for general Markov processes the GS entropy relative to the stationary measure,
\begin{align}
S_{\rm G}(\mu|\mu_{\rm s})=-\int_\Gamma\mu(X,t)\log\left(\frac{\mu(X,t)}{\mu_{\rm s}(X)}\right)\d X= S_{\rm G}(\mu) + \int \mu\log\mu_{\rm s}\d X\quad,\label{eq:reldec}
\end{align}
is monotone non decreasing \cite{doob1953stochastic}. We thus always have
\begin{align}
\frac{\d }{\d t}S_{\rm G}(\mu)+\frac{\d }{\d t}\mu\log\mu_{\rm s}\ge0\quad,\label{eq:condsec}
\end{align}
irrespective of whether the stochasticity comes from thermal reservoirs or not. This time derivative coincides in our case with $\sigma_{\rm G}$ when the system is in contact with only one reservoir and $\mu_{\rm s}\sim\exp[-\beta_\alpha H]$. When the system is in contact with several reservoirs then in addition to (\ref{eq:reldec}) we also have $\frac{\d }{\d t}S_{\rm G}(\mu)+\frac{\d }{\d t}\sum_\alpha\mu\log\mu_{\alpha}\ge0$. The positivity of $\sigma_{\rm G}$ is thus simply a consequence of (\ref{eq:condsec}) and the detailed balance condition for each reservoir which gives (\ref{eq:detbal}). Equation~(\ref{eq:condsec}) would thus hold whatever the stationary states, $\mu_\alpha$, of the system in contact with only one reservoir.


The relationship between $S_{\rm G}$ and $S_{\rm B}$ for open systems is an interesting question. It may be considered in the following example in which a macroscopic system is in contact with a single reservoir, e.g. a metal ball of radius 10 cm immersed in a large tub of water.
Consider the case when at $t=0$ we have $\mu(X,0) = \mu_{\beta_0} (X)$, i.e. the system is in equilibrium with a reservoir at the temperature $T_0$.  At $t=0$ the system is suddenly coupled to a thermal reservoir at temperature $T_f$, with which it comes to equilibrium as $t\to\infty$.
We thus have $S_{\rm G}(0)=S_{\rm B}(0)$ and $S_{\rm G}(\infty)=S_{\rm B}(\infty)$, but what about the times in between? Under the very reasonable assumption of LTE, say $T_0=50^{\rm o}{\rm C}$ and $T_f=30^{\rm o}{\rm C}$, $S_{\rm B}(t)$ can be computed for all $t>t_0$ from the heat equation, but what about $S_{\rm G}(t)$? Does it agree with $S_{\rm B}(t)$ to leading order? Or do we have $S_{\rm G}(t)<S_{\rm B}(t)$ to leading order for some values of $t$? We do not know.

A similar question can be asked  when the system is in contact with two (or more) reservoirs at different temperatures on its surface. We expect that if the system is macroscopic and chaotic, i.e. it satisfies Fourier's law, then the energy and density profile in the stationary state computed as an average over $\mu_{\rm s}$ will be that corresponding to LTE. The quantity $\sigma_{\rm G}$ will then be given by 
\begin{align}\label{eq:sigG}
\sigma_{\rm G}=-\sum_\alpha J_\alpha/T_\alpha\quad,
\end{align}
since $\d S_{\rm G}(\mu_{\rm s})/\d t=0$, in (\ref{eq:entflux}). As long as the first integral in (\ref{eq:boltentbal}) can be identified with (\ref{eq:sigG}), there will be a similar expression for the hydrodynamic entropy production $\sigma_{\rm B}$ in (\ref{eq:boltentbal}) and (\ref{eq:boltepr}). This raises the following question: to what extent does the stationary measure $\mu_{\rm s}$ for a ``chaotic'' system correspond to a LTE state when the only stochasticity is the one at the surface induced by the reservoirs. In other words, does $S_{\rm G}(\mu_{\rm s})=S_{\rm B}$ in this scenario? For the particular case where there are also bulk stochastic interactions which satisfy detailed balance this has been proven  \cite{derrida2007entropy, bonetto2004fourier}. However, for the more general case in which the bulk dynamics is Hamiltonian this remains an open question. 

In fact, it is not true that $S_{\rm G}(\mu_{\rm s})=S_{\rm B}$ when the  reservoirs at the surface are dissipative but deterministic, see \cite{chernov1997stationary}. There, it is considered a NESS produced by driving the system via deterministic non Hamiltonian forces of the type used in Gaussian thermostats at the surface. These yield a NESS, $\mu_{\rm s}$, which is singular with respect to Liouville measure. Its Gibbs entropy, $S_{\rm G}(\mu_{\rm s})$, is thus equal to $-\infty$. On the other hand molecular dynamics simulations of this model show that it is in LTE, corresponding to shear flow, as far as thermodynamic quantities are concerned. Whether such a situation can also occur when the NESS is produced by stochastic thermal reservoirs is an open question. This problem is also discussed in section 6 of \cite{lebowitz1999gallavotti} for different kinds of reservoirs.
 
\section{Small system in contact with a thermal reservoir}
Isolated small classical systems, such as a few particles in a box, and systems with only a few relevant degrees of freedom, such as the center of mass motion of a massive pendulum or the moon, are not thought to have any thermodynamic functions, such as entropy, associated with them. For this reason the second law is, as noted by Maxwell, constantly being violated in small systems \cite{maxwell}.  It is certainly no great surprise if an isolated box containing 10 Argon atoms is frequently seen to have 8 or more particles in the right half of the box.  Such a percentage of particles on one side of an isolated system would certainly be a violation of the second law if the system consisted of $10^{20}$ or more particles.  Just how large does the system have to be to rule out ``ever'' seeing such a violation in an isolated system of $N$ particles during a period of 100 years depends (strongly) on the nature of the interaction between the particles, shape of the container, initial state, and on whether we are considering classical or quantum dynamics. Leaving quantum systems for a separate consideration, we will analyze now what happens to a small classical system in contact with a thermal reservoir.

The Hamiltonian of such a small system in contact with a thermal reservoir is given by
\begin{align}
H_{\rm tot} = H_{\rm sys}(X)+H_{\rm r}(Y)+V(X,Y)\quad,
\end{align}
where $X$ describes the relevant part of the microstate of the small  system with Hamiltonian $H_{\rm sys}$, $Y$ that of the reservoir with Hamiltonian $H_{\rm r}$, and $V$ is the interaction between system and reservoir. If the total system is in equilibrium and is described by a microcanonical or canonical ensemble at a temperature $\beta^{-1}$, this induces a probability density for the system $\tilde{\mu}(X)\sim\int \mu_\beta(X,Y)\d Y\sim \exp[-\beta(H_{\rm sys}(X)+\tilde{V}(X,\beta))]$. Note that $\tilde{V}$ will generally be determined by both, $H_{\rm r}$ and $V$, and can depend on $\beta$. This has to be taken into account when one considers, for example, the collapse transition of a polymer in a solvent \cite{jarzynski2017stochastic, lebowitz2015equilibrium}.

We note here that $\tilde{\mu}(X)$ no longer gives almost sure predictions about the properties of the small system. We will presumably however get the same $\tilde{\mu}(X)$ when the size of the reservoir is very large for all different Gibbs ensembles describing the total system. Nevertheless, it is not clear how meaningful it is to assign thermodynamic functions to the small system based on $\tilde{\mu}(X)$: see discussion in \cite{jarzynski2017stochastic}. We shall focus here on cases where the interaction $V(X,Y)$ can be taken to be of the impulsive type, as in section 3, where $\tilde{V}(X)$ can be taken to be essentially independent of $X$ and set equal to zero. The paradigm of such a system is a Brownian particle [BP] immersed in an equilibrium fluid at temperature $T$.  The only relevant degree of freedom for such a (spherical) particle is the location of its center of mass. The phase space $\Gamma$ of the system is thus six dimensional, $X=(\vec{r},\vec{v})$, where we have set the mass of the BP equal to unity so that $\vec{p}=\vec{v}$. Treating the fluid (approximately) as an infinite thermal reservoir one obtains a stochastic Liouville equation of the form of Eq.~(\ref{eq:liouville}) for $\mu(\vec{r}, \vec{v}, t)$ with $H (\vec{r},\vec{v}) = \frac12  |\vec{v}|^2 + U (\vec{r})$, where $U(\vec{r})$ is an external potential which varies slowly on the microscopic spatial scale.

For a sufficiently idealized fluid in thermal equilibrium at temperature $\beta^{-1}$ one can obtain, in an appropriate limit, a Fokker-Planck equation for the time evolution of the probability density of the BP $\mu(\vec{r}, \vec{v}, t)$
\begin{align}
\frac{\partial\mu}{\partial t} + \vec{v}\cdot \frac{\partial \mu}{\partial \vec{r}} - \frac{\partial U}{\partial \vec{r}}\cdot \frac{\partial \mu}{\partial \vec{v}} = \xi \frac{\partial}{\partial \vec{v}} \cdot\left[\mu_\beta \frac{\partial}{\partial \vec{v}} \left(\mu / \mu_\beta\right)\right]\quad,\label{eq:langevin}
\end{align}
where $\mu_\beta=Z^{-1}\exp \left\{ - \beta \left[\frac{1}{2} |\vec{v}|^2 + U (\vec{r})\right]\right\}$, c.f. \cite{durr1981mechanical}. Thus within the approximate, but physically appropriate, scheme Eq.~(\ref{eq:langevin}) treats the fluid as an infinite thermal reservoir which exerts a stochastically stationary, delta-time correlated, Gaussian force on the particle. The particle distribution then evolves towards its stationary value $\mu_\beta$  on a time scale $T/\xi$.


For the Fokker-Planck equation (\ref{eq:langevin}), one can formally follow the approach of section~\ref{sec:timsgibbs} \cite{bergmann1955new, lebowitz1957irreversible} to calculate the corresponding change in the Gibbs-Shannon entropy production. The result is given by
\begin{align}
\sigma_{\rm G} &= \xi\int \mu (\vec{r}, \vec{v}, t)\left|\frac {\partial }{\partial \vec{v}} \log \nu \right|^2 \d\vec{r} \d\vec{v}= \frac{\d S_{\rm G}}{\d t} - J/T\ge0 \quad,\label{eq:lte}
\end{align}
where {$\nu= \mu/\mu_{\beta}$ as in Eq.~(\ref{eq:nu})} and $J$ is the average energy flux from the fluid to the Brownian particle $J=\frac{\d}{\d t}\int (\frac12 |\vec{v}|^2+U) \mu\d\vec{r}\d\vec{v}$. One can take further limits when the Fokker-Planck equation becomes a Langevin equation but we shall not go into that here \cite{risken1996fokker}.

Equations (\ref{eq:langevin}) and (\ref{eq:lte}) and their analogues play a central role in ``stochastic thermodynamics'' where $S_{\rm G}(\mu)$ is generally taken for granted to represent a thermodynamic entropy and thus (\ref{eq:lte}) is considered to be an expression of the second law. There are in fact, as already noted, recent experiments which give some support to this interpretation \cite{gavrilov2017direct}. The question therefore naturally arises of why this should be true for small systems in contact with thermal reservoirs when, as argued above, this is not the case for isolated systems and may not be true for MS in contact with reservoirs at their surfaces. 


\section{The Brownian gas}
\label{sec:bg}
We shall now attempt to justify the identification of $S_{\rm G}(\mu)$ of a nano-particle in contact with a thermal reservoir, such as a BP in a fluid, with a thermodynamic entropy. Consider a dilute gas of $N$ of such BP, $N\gg1$, and call it a Brownian gas [BG]. The gas is so dilute that interactions between the BP are negligible. This BG is a macroscopic system in contact with a thermal reservoir not just at its boundaries, but ``everywhere''. Let $\gamma$ be the $6$ dimensional phase space of the Brownian particle (in the older literature $\gamma$ is called the $\mu-$space, where $\mu$ stands for molecule). Then, the phase space $\Gamma$ of the BG will have $6N$ dimensions. 

The (``meso'') macrostate of the Brownian gas is given by specifying the number of Brownian particles in each region $\d\vec{r}\d\vec{v}$ of the $6$ dimensional $\gamma$ space, to be {$Nf(\vec{r},\vec{v})\d\vec{r}\d\vec{v}$}. This corresponds to a region $\Gamma_f$ in the $6N$ dimensional phase space $\Gamma$. The $\log$ of the Liouville volume $|\Gamma_f|$ is given, up to constants, by
\footnote{The derivation of Eq.~(\ref{eq:phase}), due to Boltzmann, is straightforward. Divide the $\gamma-$space into regions $\Delta_\alpha$, with $\alpha=1,\ldots,M$, and let $N_\alpha$ be the number of particles in $\Delta_\alpha$. Then, one has that $|\Gamma_f|\sim\prod\frac{|\Delta_\alpha|^{N_\alpha}}{N_\alpha!}$. Using Stirling's formula, one obtains Equation~(\ref{eq:phase}), see \cite{goldstein2004boltzmann} for details.}
\begin{align}
\log|\Gamma_f| = -N\int f(\vec{r},\vec{v})\log f(\vec{r},\vec{v})\d\vec{r}\d\vec{v}+ N\quad.\label{eq:phase}
\end{align} 
The Boltzmann entropy, $S_{\rm B}(f)$, of this meso state is then given by (\ref{eq:phase}), whose right hand side coincides up to a constant term with $NS_{\rm G}(\mu)$. This is so even though the physical interpretations of $NS_{\rm G}(\mu)$ and $S_{\rm B}(f)$ are quite different.

The entropy production in a Brownian gas plus fluid is given by 
\begin{align}
\sigma_{\rm B} = -N\frac{\d }{\d t}\int f\log f\d\vec{r}\d\vec{v} - J_N/T\ge0\quad,\label{eq:boltzmannepr}
\end{align}
where  $J_N$ is the flux of energy from the fluid to the Brownian gas, and $-J_N/T$ is the rate of entropy change of the fluid at temperature $T$. The right side of equation (\ref{eq:boltzmannepr}) is just $N$ times  the right hand side of equation (\ref{eq:lte}) if one identifies $J_N=NJ$. 

We remark here that Boltzmann's famous $\mathcal{H}-$theorem shows the monotone increase of $-\int f\log f\d \vec{r}\d\vec{v}$ for an isolated dilute gas evolving in time according to the Boltzmann equation. Boltzmann interpreted this as a microscopic derivation of the second law for $S_{\rm B}(\{ f\})$ and says \cite{boltzmann2017vorlesungen}: ``we have thus succeeded in defining entropy for a system not in equilibrium''.

An important observation now is that, unlike an isolated gas, where some interaction between the particles is essential to make the system satisfy the second law (rather than behaving like an ideal gas), the Brownian gas gets thermalized via its interaction with the fluid. Hence the behavior of a single Brownian particle averaged over many trials will be the same as that of a Brownian gas. It is therefore meaningful to consider the Gibbs-Shannon entropy of a single Brownian particle as having a thermodynamic meaning, i.e. being equal to that of a Brownian gas divided by the number of particles. This should be true both when the Brownian gas is in global equilibrium, or in a meso (macro) state described by $f(\vec{r},\vec{v})$.

The above considerations will hold also in the case when the Brownian particle is acted on by a time dependent external potential $U(\vec{r}, t)$. The Brownian gas will behave like a MS on which work is being done. In particular when $U(\vec{r},t)$ varies sufficiently slowly in time compared to the time it takes the Brownian particle to relax to equilibrium with $U(\vec{r}, t)$, then the entropy will change adiabatically and the right hand side of (\ref{eq:lte}) will be an equality. The behavior of a single Brownian particle will then be similar to that of the Brownian gas, with vanishing fluctuations. This is  what is observed in the experiments in \cite{gavrilov2017direct} which we discuss next. 

\section{Experiments on a Brownian particle}

An idealized version of the experiment in \cite{gavrilov2017direct} is as follows: The thermal reservoir fluid occupies a volume $V$ which is divided into regions $V_1$ and $V_2.$ At $t=0$ the BP is in equilibrium with the fluid in $V_1$ and a confining (infinite) external potential $U_0(\vec{r})$ which excludes it from $V_2$. At $t=0$, $U_0(\vec{r})$ is changed to $U_1(\vec{r})$ without any work being done, e.g. one suddenly removes the infinite potential confining the particle to $V_1$. One then waits until time $t_1$ for the particle to come to equilibrium with the fluid at the new potential $U_1(\vec{r})$, e.g. no confining potential. One then changes $U_1$ to $U_0$ by gradually raising the height of the potential in $V_2$ over a time interval $\tau$. During this time one does work $W(\tau)$ on the particle.

We make the time variation of $U(\vec{r},t)$ during $\tau$ very slow compared to the relaxation time of the particle to its equilibrium distribution. Hence during the time interval $\tau$ of a given realization the probability of the particle being in position $\vec{r}$ with velocity $\vec{v}$ varies in a quasistatic way. From a thermodynamic point of view the macro (meso) state of the corresponding Brownian gas is given up to a factor $N$ by
\begin{align}
\mu_{\beta}(\vec{r},\vec{v},t)=\frac{1}{Z(t)}\exp\left(-\beta\left[\frac{1}{2}|\vec{v}|^2+U(\vec{r},t)\right]\right)
\end{align}
with $U(\vec{r},t_1)=U_1(\vec{r})$ and $U(\vec{r},t_1+\tau)=U(\vec{r},0)=U_0(\vec{r})$.

We can now use standard thermodynamics to calculate the work done by an external agent that slowly manipulates the potential $U(\vec{r},t)$. The  work done per unit time is $\dot w(t)=- \vec{v}\cdot{\partial U}/{\partial \vec{r}}|_{\vec{r}(t)}$, where $\vec{r}(t)$ is the position of the BP at time $t$. The total work done over the duration of the period $(0,t_1+\tau)$ is then different from zero only during the interval $(t_1,t_1+\tau)$, and is given by
\begin{align}
W(\tau) = \int_{t_1}^{t_1+\tau} \dot{w}(t) \d t=\int_{t_1}^{t_1+\tau} \frac{\partial U(\vec{r},t)}{\partial t} \d t\quad,
\end{align}
where it is important for the equality that the potential is the same at the beginning and at the end of the protocol \cite{vilar2008failure, peliti2008work}. We can now relate the average total work $\langle W(\tau)\rangle$, where the average is taken with respect to $\mu_\beta(\vec{r},\vec{v},t)$, with the change of $\log Z$ during the period $\tau$. More precisely
\begin{align}
\langle W(\tau)\rangle &= \int\left(\int_{t_1}^{t_1+\tau}\frac{\partial U(\vec{r},t)}{\partial t}\mu_\beta(\vec{r},\vec{v},t)\d t\right)\d \vec{r}\d\vec{v}= -\int\left(\int_{t_1}^{t_1+\tau}\frac{1}{\beta Z(t)}\frac{\partial}{\partial t}\left[ \e^{-\beta(\frac12|\vec{v}|^2+U (\vec{r},t))} \right] \right)\d\vec{r}\d\vec{v}\nonumber\\
&= T\log[Z(t_1)/Z(t_1+\tau)]\quad.\label{eq:workent}
\end{align}
To interpret this work as a change in Gibbs-Shannon entropy we can integrate by parts in the definition of $\langle W(\tau)\rangle$. This gives
\begin{align}\label{eq:workent}
\langle W(\tau)\rangle = E(t_1+\tau)-E(t_1)-T[S_{\rm G}(t_1+\tau)-S_{\rm G}(t_1)]\quad,
\end{align}
where we have defined the average energy as $E(t)=\langle \frac12|\vec{v}|^2+U (\vec{r},t)\rangle$. In the actual experiment $E(t_1)=E(0)=E(t_1+\tau)$. Hence measuring $\langle W(\tau)\rangle$, which was shown experimentally to have very little variance for large $\tau$, lets one measure $S_{\rm G}(\mu_\beta(t_1))-S_{\rm G}(\mu_\beta(t_1+\tau))$ for different external potentials $U(\vec{r},t)$, e.g. for different confining volumes of $U_0(\vec{r})$.

Using the BG interpretation, the quantities in equation~(\ref{eq:workent}) can all be interpreted as macroscopic quantities divided by the number of particles. Within this interpretation $S_{\rm G}$ coincides with the hydrodynamic entropy which changes in an irreversible way. The entropy of the heat bath (here the fluid) has increased during the cycle from $0$ to $t_1+\tau$ by $\langle W(\tau)\rangle/T$. When $\tau$ is not so large so that one can not assume instantaneous equilibrium of the BG there will be extra entropy production in the BG during this period. This work, obtained by averaging $W(\tau)$ over repetitions of the experiment, was indeed found to be greater than the right hand side of  (\ref{eq:workent}). In the analysis of the experiment and in stochastic thermodynamics one goes beyond the simple equality or inequality of (\ref{eq:workent}). One actually computes the distribution of $W(\tau)$. We shall not go into that here. We thus conclude that the interpretation of $S_{\rm G}(\mu)$ as the thermodynamic Boltzmann entropy per particle of a Brownian gas is consistent.

\section{Concluding remarks}

\label{sec:last}

The point of view taken in this note is that 
the entropy of a physical system should be a property of the state of the {\it individual} system and thus it should be possible to define the entropy of a system without referring to any ensembles, see \cite{lebowitz2007time, goldstein2004boltzmann}.  For a classical system the most detailed description of the physical state of the system is that given by its microstate $X\in\Gamma$.  Thus any physical entropy $S$ is a function of $X$. This is the case for the Boltzmann entropy $S_{\rm B}(X)$ of a macroscopic system in a well defined macrostate $M$ (for which $S_{\rm B}$ is in fact the same for all $X\in \Gamma(M)$). Going beyond the  hydrodynamic entropy (\ref{ent:hydro}), appropriate only for systems in local thermal equilibrium, $S_{\rm B}$ can be extended to dilute gases not in local thermal equilibrium.  For these the macro (meso) state $M$ is specified by the empirical distribution $f(\vec{r}, \vec{v}, t)$, which is of course determined by $X$, see \cite{goldstein2004boltzmann}.  In contrast the Gibbs-Shannon entropy (of a measure) not only fails to be determined by the microstate of the system -- it also fails to change in time for an isolated system, large or small, even for a large isolated system that is undergoing (internally) dissipative relaxation and thus producing thermodynamic entropy.  Of course the Gibbs-Shannon entropy of the microcanonical ensemble $\mu_{\rm m}$ is meaningful for an isolated macroscopic system in global thermal equilibrium, where it coincides with $S_{\rm B}$ to leading order in the size of the system. 

The existence of a useful general notion of entropy for an isolated nanosystem  is not so clear. Such systems have been studied theoretically and experimentally for quantum systems \cite{kaufman2016quantum, goldstein2015thermal}, which we have not discussed here.  In fact our main concern here has been with the significance of $S_{\rm G}$ for systems large and small in contact with heat baths.  We have not resolved this issue for macroscopic systems, but have given a possible, to us plausible, answer for the case of a nanosystem studied experimentally in \cite{gavrilov2017direct}.
\newline
\newline

{\bf Acknowledgements:}  We thank John Bechhoefer, Rafa\"el Chetrite, Stanislas Leibler, Eugene Speer and Bingkan Xue for fruitful discussions. The work of JLL was supported by an AFOSR grant FA9550-16-1-0037. The work of PS  has  been  partly  supported  by  grants  from  the  Simons Foundation to Stanislas Leibler through The Rockefeller University (Grant 345430) and the Institute for Advanced Study (Grant 345801). DAH, JLL, and PS thank the Institute for Advanced Study for its hospitality during the elaboration of this work.


\begin{thebibliography}{10}

\bibitem{klein1973development}
M.~J. Klein, ``The development of Boltzmann's statistical ideas,''  E.~G.~D.~Cohen \& W.~Thirring, eds. {\em The Boltzmann equation. Theory and application}, Springer-Verlag, Berlin, pp.~53--106, 1973.

\bibitem{penrose2005foundations}
O.~Penrose, {\em Foundations of statistical mechanics: a deductive treatment}.
\newblock Courier Corporation, 2005.

\bibitem{lebowitz2007time}
J.~L. Lebowitz, ``From time-symmetric microscopic dynamics to time-asymmetric
  macroscopic behavior: An overview,'' {\em Boltzmann's Legacy}, pp.~63--88,
  2007.

\bibitem{goldstein2004boltzmann}
S.~Goldstein and J.~L. Lebowitz, ``On the (Boltzmann) entropy of
  non-equilibrium systems,'' {\em Physica D: Nonlinear Phenomena}, vol.~193,
  no.~1, pp.~53--66, 2004.

\bibitem{kipnis2013scaling}
C.~Kipnis and C.~Landim, {\em Scaling limits of interacting particle systems},
  vol.~320.
\newblock Springer Science \& Business Media, 2013.

\bibitem{giacomin1998deterministic}
G.~Giacomin, J.~L. Lebowitz, and E.~Presutti, ``Deterministic and stochastic
  hydrodynamic equations arising from simple microscopic model systems,'' {\em
  Mathematical Surveys and Monographs}, vol.~64, pp.~107--152, 1998.

\bibitem{lebowitz1988microscopic}
J.~L. Lebowitz, E.~Presutti, and H.~Spohn, ``Microscopic models of hydrodynamic
  behavior,'' {\em Journal of Statistical Physics}, vol.~51, no.~5,
  pp.~841--862, 1988.

\bibitem{gavrilov2017direct}
M.~Gavrilov, R.~Ch{\'e}trite, and J.~Bechhoefer, ``Direct measurement of weakly
  nonequilibrium system entropy is consistent with Gibbs--Shannon form,'' {\em
  Proceedings of the National Academy of Sciences}, vol.~114, no.~42,
  pp.~11097--11102, 2017.

\bibitem{ciliberto2017experiments}
S.~Ciliberto, ``Experiments in stochastic thermodynamics: Short history and
  perspectives,'' {\em Physical Review X}, vol.~7, no.~2, p.~021051, 2017.

\bibitem{kaufman2016quantum}
A.~M. Kaufman, M.~E. Tai, A.~Lukin, M.~Rispoli, R.~Schittko, P.~M. Preiss, and
  M.~Greiner, ``Quantum thermalization through entanglement in an isolated
  many-body system,'' {\em Science}, vol.~353, no.~6301, pp.~794--800, 2016.

\bibitem{brush1967science}
S.~G. Brush, {\em Science and culture in the nineteenth century: thermodynamics
  and history}.
\newblock Univerity of Texas, 1967.

\bibitem{callen1998thermodynamics}
H.~B. Callen, {\em Thermodynamics and an Introduction to Thermostatistics}.
\newblock John Wiley and sons, 1998.

\bibitem{de2013non}
S.~R. De~Groot and P.~Mazur, {\em Non-equilibrium thermodynamics}.
\newblock Courier Corporation, 2013.

\bibitem{shannon1948mathematical}
C.~Shannon, ``A mathematical theory of communication,'' {\em The Bell System
  Technical Journal}, vol.~27, no.~3, pp.~379--423, 1948.

\bibitem{ehrenfest2002conceptual}
P.~Ehrenfest and T.~Ehrenfest, {\em The conceptual foundations of the
  statistical approach in mechanics}.
\newblock Courier Corporation, 2002.

\bibitem{bergmann1955new}
P.~G. Bergmann and J.~L. Lebowitz, ``New approach to nonequilibrium
  processes,'' {\em Physical Review}, vol.~99, no.~2, p.~578, 1955.

\bibitem{lebowitz1957irreversible}
J.~L. Lebowitz and P.~G. Bergmann, ``Irreversible gibbsian ensembles,'' {\em
  Annals of Physics}, vol.~1, no.~1, pp.~1--23, 1957.

\bibitem{doob1953stochastic}
J.~L. Doob, {\em Stochastic processes}.
\newblock Wiley, New York, 1953.

\bibitem{derrida2007entropy}
B.~Derrida, J.~Lebowitz, and E.~Speer, ``Entropy of open lattice systems,''
  {\em Journal of Statistical Physics}, vol.~126, no.~4-5, pp.~1083--1108,
  2007.

\bibitem{bonetto2004fourier}
F.~Bonetto, J.~L. Lebowitz, and J.~Lukkarinen, ``Fourier's law for a harmonic
  crystal with self-consistent stochastic reservoirs,'' {\em Journal of
  statistical physics}, vol.~116, no.~1, pp.~783--813, 2004.

\bibitem{chernov1997stationary}
N.~Chernov and J.~L. Lebowitz, ``Stationary nonequilibrium states in
  boundary-driven hamiltonian systems: shear flow,'' {\em Journal of
  statistical physics}, vol.~86, no.~5, pp.~953--990, 1997.

\bibitem{lebowitz1999gallavotti}
J.~L. Lebowitz and H.~Spohn, ``A Gallavotti--Cohen-type symmetry in the large
  deviation functional for stochastic dynamics,'' {\em Journal of Statistical
  Physics}, vol.~95, no.~1, pp.~333--365, 1999.

\bibitem{maxwell}
J.~C. Maxwell, {\em ``Theory of Heat'', p.308: Tait's Thermodynamics, Nature
  17, 257 (1878).} Quoted in M. J. Klein {\em``The development of Boltzmann's
  statistical ideas''}.
\newblock see ref. [1].

\bibitem{jarzynski2017stochastic}
C.~Jarzynski, ``Stochastic and macroscopic thermodynamics of strongly coupled
  systems,'' {\em Physical Review X}, vol.~7, no.~1, p.~011008, 2017.

\bibitem{lebowitz2015equilibrium}
J.~L. Lebowitz and L.~Pastur, ``On the equilibrium state of a small system with
  random matrix coupling to its environment,'' {\em Journal of Physics A:
  Mathematical and Theoretical}, vol.~48, no.~26, p.~265201, 2015.

\bibitem{durr1981mechanical}
D.~D{\"u}rr, S.~Goldstein, and J.~Lebowitz, ``A mechanical model of brownian
  motion,'' {\em Communications in Mathematical Physics}, vol.~78, no.~4,
  pp.~507--530, 1981.

\bibitem{risken1996fokker}
H.~Risken, ``Fokker-planck equation,'' in {\em The Fokker-Planck Equation},
  pp.~63--95, Springer, 1996.

\bibitem{boltzmann2017vorlesungen}
L.~Boltzmann, {\em Vorlesungen {\"u}ber Gastheorie: 2. Teil, Leipzig: Barth,
  1896, 1898}.
\newblock This book has been translated into English by S.G. Brush, Lectures on
  Gas Theory, Cambridge University Press, London, (1964).

\bibitem{vilar2008failure}
J.~M. Vilar and J.~M. Rubi, ``Failure of the work-hamiltonian connection for
  free-energy calculations,'' {\em Physical review letters}, vol.~100, no.~2,
  p.~020601, 2008.

\bibitem{peliti2008work}
L.~Peliti, ``On the work--hamiltonian connection in manipulated systems,'' {\em
  Journal of Statistical Mechanics: Theory and Experiment}, vol.~2008, no.~05,
  p.~P05002, 2008.

\bibitem{goldstein2015thermal}
S.~Goldstein, D.~A. Huse, J.~L. Lebowitz, and R.~Tumulka, ``Thermal equilibrium
  of a macroscopic quantum system in a pure state,'' {\em Physical review
  letters}, vol.~115, no.~10, p.~100402, 2015.

\end{thebibliography}

\end{document}